\documentclass[prl, reprint,aps, superscriptaddress,noeprint]{revtex4-1}
\usepackage{graphicx}
\usepackage{textcomp}
\usepackage{amsmath}
\usepackage{amsfonts}
\usepackage{mathtools}
\usepackage{color}
\usepackage{mathrsfs}
\usepackage{scrextend}

\newcommand{\fig}{Fig. }
\newcommand{\eqn}{Eqn. }

\begin{document}


\title{Dynamics of Reaction-Diffusion Oscillators in  Star Networks}



\author{Michael M. Norton}
\thanks{authors contributed equally to this work}
\affiliation{Physics Department, Brandeis University, Waltham, Massachusetts 02453}

\author{Nathan Tompkins}
\thanks{authors contributed equally to this work}
\affiliation{Physics Department, Brandeis University, Waltham, Massachusetts 02453}
\affiliation{Physics Department, Wabash University, Crawfordsville, Indiana 47933}

\author{Baptiste Blanc}
\affiliation{Physics Department, Brandeis University, Waltham, Massachusetts 02453}

\author{Matthew Carl Cambria}
\affiliation{Physics Department, Brandeis University, Waltham, Massachusetts 02453}

\author{Jesse Held}
\affiliation{Physics Department, Brandeis University, Waltham, Massachusetts 02453}

\author{Seth Fraden}
\email{fraden@brandeis.edu}
\affiliation{Physics Department, Brandeis University, Waltham, Massachusetts 02453}


\date{\today}

\begin{abstract}
We experimentally and theoretically investigate the dynamics of inhibitory coupled self-driven oscillators on a star network in which a single central hub node is connected to $k$ peripheral arm nodes. The system consists of water-in-oil Belousov-Zhabotinsky $\sim$100$\mu$m emulsion drops contained in storage wells etched in silicon wafers. We observed three dynamical attractors by varying the number of arms in the star graph and the coupling strength; (\textit{i}) \textit{unlocked}; uncorrelated phase shifts between all oscillators, (\textit{ii}) \textit{locked}; arm-hubs synchronized in-phase with a $k$-dependent phase shift between the arm and central hub, and (\textit{iii})  \textit{center silent};  central hub stopped oscillating and the arm-hubs oscillated without synchrony. We compare experiment to theory. For case (\textit{ii}), we identified a logarithmic dependence of the phase shift on star degree, and were able to discriminate between contributions to the phase shift arising from star topology and oscillator chemistry.
\end{abstract}

\maketitle

Discrete networks of self-driven oscillators represent a broad class of physical systems \cite{Pecora2014b,sorrentino2016complete,Kopell:1988,Strogatz:Sync,reimann2017cliques, Barabasi:Applause,Pantaleone:2002,Dorfler:2013, Synchronization,Strogatz:Chaos,Kuramoto:Waves}. Idealized models have the virtue of being simple enough to explain  network dynamics, but raise the question of whether such models can quantitatively characterize natural systems.  Addressing this question has motivated the development of controlled, model experimental systems that operate with varying degrees of autonomy \cite{jia2012spontaneously, Kiss:2013, Kiss:2014, Horvath:2015, NingPCCP:2014, Ning:2014, Tompkins:2015, Litschel2017, Blaha2019}. A particularly important goal of oscillator network research is to identify  how (i) individual oscillator's dynamics, (ii) network topology and (iii) coupling type conspire to create emergent spatio-temporal patterns. 

\begin{figure}
\begin{center}
\includegraphics[width=\columnwidth]{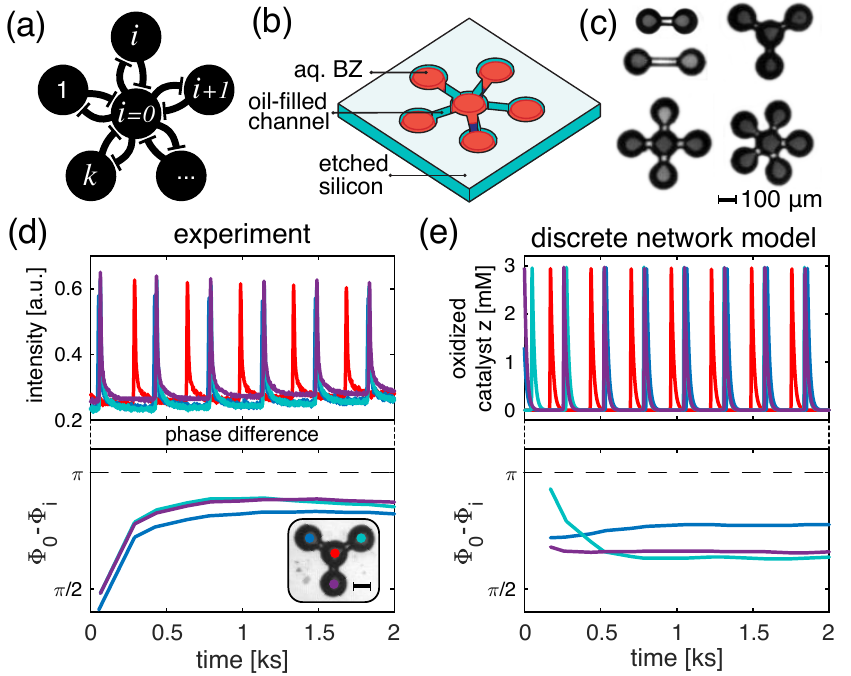}
\caption{\textbf{(a)} Schematic of inhibitory-coupled star network of degree $k$. \textbf{(b)} Schematic of experimental setup: BZ aqueous drops are stored in circular wells etched in a silicon wafer and  separated by fluorinated oil. \textbf{(c)} Reflection microscopy images of loaded star networks with different numbers of arms and arm lengths; movies S1-S8, \textbf{(d)} and \textbf{(e)} time trace of droplet intensity and corresponding evolution of arm node phase relative to hub node showing phase-locking for experiment (movie S2) and point model simulation (\eqn \ref{eqn:DRDN}).}
\label{fig:Dynamics}
\end{center}
\end{figure}

To address this goal, we experimentally study and theoretically model a minimally complex microfluidically-assembled discrete network of Belousov-Zhabotinsky (BZ) chemical oscillators. Notably, our BZ system uses  natural physical-chemical processes to produce the oscillator dynamics, network topology and coupling, in contrast to systems in which the oscillators, coupling, or both are mediated through electronic hardware \cite{Kinoshita:2006,Kiss2006,Pecora2014b,sorrentino2016complete,Totz2018,Blaha2019}.  Thus, our system also demonstrates the potential of engineering stand-alone soft materials to exhibit self-organized spatiotemporal patterns. An application of such materials is to make autonomous, soft robots that run purely on chemicals like living organisms, rather than powered by motors and controlled by computers. In this scenario, the BZ network would function as a central pattern generator found in the nervous system of many animals \cite{Litschel2017}. The BZ reaction requires only a few reagents and networks can be fabricated and parallelized through microfluidic techniques \cite{Tompkins:2015}. The theoretical foundation of BZ chemistry is strong, enabling near-quantitative modeling of both isolated and coupled oscillators. We examine inhibitory coupling, which promotes symmetry-breaking phenomena by preferring anti-phase synchrony between neighbors \cite{Epstein:2009, Vanag2011, Tompkins:2014, NingPCCP:2014, Guzowski2015a, torbensen2017chemical, Litschel2017}. We arrange the cells in a network with star topology, consisting of a central hub node connected by $k$-arms to other nodes, as illustrated in \fig\ref{fig:Dynamics}. Star networks have been considered in idealized theoretical and experimental studies \cite{Pecora1998b, Bergner2012} and are a naturally occurring motif in neural networks that perform cognitive \cite{Bonifazi2009,VandenHeuvel2013,Bertolero2018} and sensorial functions \cite{Khaledi-Nasab2018}.


We probe the dynamical states of the system as a function of inter-nodal coupling strength and star degree, thereby focusing on the impact of physical changes to the star network while taking the BZ chemistry as a known in our experiments and models. We compare observations to two levels of theory: a discrete reaction-diffusion model with Vanag-Epstein \cite{Epstein:2009} chemical dynamics taking place at each node, and a phase model constructed from it that we use to examine the topology-dependence of locking angles. Strikingly, we find that the locking angle between arm and hub nodes has a logarithmic dependence on the number of arm nodes in the limit of weak coupling.  The prefactor for this dependence depends only on the interaction function, which can be numerically derived for any oscillator from a model or an experimentally acquired phase response curve \cite{Stankovski2017}. Our result, general to all star networks,  disentangles oscillator and coupling physics from topological effects by compactly showing how each come together to produce topology-dependent phase locking.

\begin{figure}
\begin{center}
\includegraphics[width=\columnwidth]{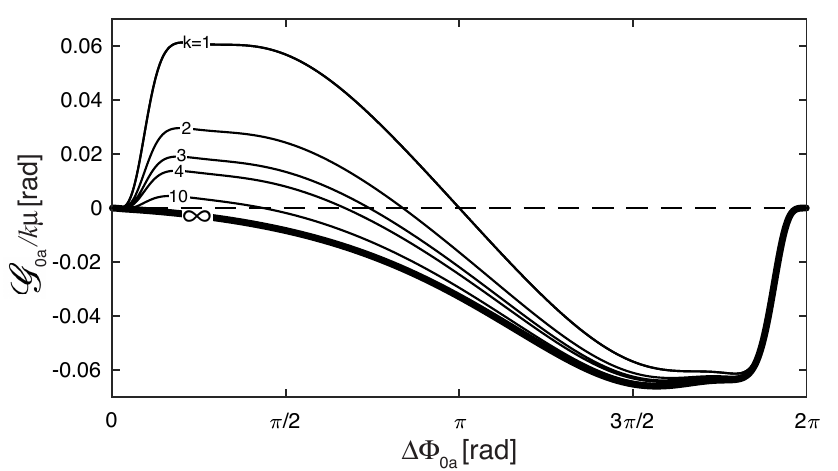}
\caption{$\mathscr{H}_{ij}/\mu$ (thick line) describes the change in phase, $\Phi_i$, oscillator $i$ experiences due to the presence of oscillator $j$ as a function of the phase difference $\Phi_i - \Phi_j$ \cite{Note1}. Dynamics $\mathscr{G}_{0a}\left(\Phi_{0a}\right)/\left(k\mu\right)$ along the arm-synchronized manifold $\Phi_1=\Phi_2= \dots = \Phi_k$ (thin lines) for different $k$ (\eqn\ref{eqn:coarsephasemodel}). Roots correspond to fixed points. As $k$ increases, the interaction becomes increasingly directional, approaching $\mathscr{H}_{ij}$ in the large $k$ limit.}
\label{fig:phasemodel}
\end{center}
\end{figure} 

\begin{figure}
\begin{center}
\includegraphics[width=3.2in]{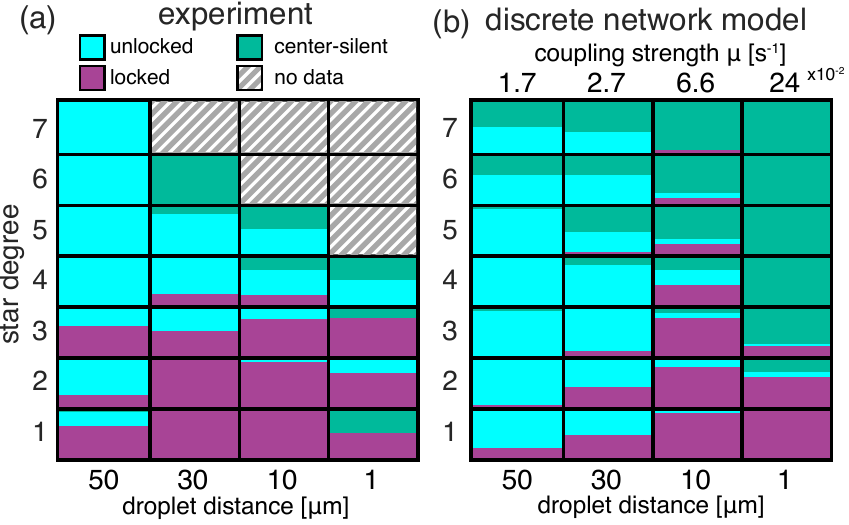}
\caption{\textbf{State Diagrams}  as a function of  arm separation (plotted from large to small lengths) and star degree for \textbf{(a)} experiment and \textbf{(b)} point model predictions, \eqn\ref{eqn:DRDN}, with 7.5\% variations in H$^+$ concentration.  Hatched region indicates where data was unobtainable due to geometric constraints. Examples of each dynamical state are shown in movies S2-S6.}
\label{fig:phasediagram}
\end{center}
\end{figure}

\textbf{Experimental System.} Surfactant stabilized emulsions of 100 micron diameter drops containing the aqueous BZ solution were generated in a fluorinated oil~\cite{Epstein:2011}.  A drop of the concentrated emulsion was pipetted onto the etched silicon wafer, a cover glass was laid on top and clamped together, thereby squeezing the BZ drops into the etched network and sealing the device. Each wafer contained hundreds of star networks. The design concept was to make a device that is simple and quick to load, intolerant to failure, and reusable. Typically 90\% of the networks fail to load correctly. However, 10 - 20 successes are enough to accumulate statistics. Both the silicon and glass are completely impermeable to all chemical species  ensuring that each arm-drop only communicates with the hub-drop~\cite{Tompkins:2015}. The cavities containing the BZ drops are connected with channels designed to be too narrow to house drops, but contain oil, and therefore function as diffusive conduits, as illustrated in \fig\ref{fig:Dynamics}B. The BZ reaction oscillates between a reduced and oxidized state of the catalyst. The duration of the oxidized, or activated state, is brief and during this interval a large amount of the inhibitor, bromine, is also generated. The inhibitory coupling between two drops is provided by bromine, which due to its low polarizability readily partitions into the oil~\cite{NingPCCP:2014}. All the chemicals, their concentrations, and conditions for producing the emulsion are described in the supplement.

\textbf{Theory and Model.} In a reaction-diffusion network consisting of $i=1\cdots N$ nodes and $m=1\cdots M$ species, the dynamics of the concentration $c_i^m$ are governed by,
\begin{equation}
\dot{c}_i^m = F_i^m\left(\mathbf{c}_i\right) + \sum_{j=1}^{N}\mu_{ij}^m \mathcal{A}_{ij} \left(c^m_j-c^m_i\right),
\label{eqn:DRDN}
\end{equation}
where $F_i^m\left(\mathbf{c}_i\right)$ models intra-nodal reactions given by the Vanag-Epstein model of BZ chemistry \cite{Epstein:2009} and the second term captures inter-nodal diffusive transport proportional to the species-dependent coefficient $\mu_{ij}^m$; full expressions are found in the supplement \cite{BZSTARsupp}. This model assumes that all the chemistry occurs at discrete points, i.e. it ignores concentration gradients within a drop. The point approximation is justified because the width of the oxidation front in the BZ reaction is larger than the size of the droplets, thus each reactor oxidizes uniformly.

Modeling the coupling as linearly proportional to the concentration difference between connected nodes is equivalent to ignoring any chemical reactions and accumulation of chemicals in the oil separating drops, which is justified when the inter-drop gap is much less than the diffusion length scale $l\sim\sqrt{D T}$, where $D$ is the diffusivity of bromine in oil and the timescale is given by the oscillation period $T$ \cite{NingPCCP:2014,Ning:2014, Tompkins:2015}. We consider gaps no greater than 60 $\mu$m. With diffusivity $D\sim\mathscr{O}\left(10^{-9}\right)$ m$^2$s$^{-1}$ and oscillation period $T\sim\mathscr{O}\left(10^{2}\right)$s,  the diffusion length scale is $l\sim$300$\mu$m; thus we are safely in the quasi-steady regime. Network connectivity is given by the adjacency matrix $\mathbf{\mathcal{A}}$ such that $\mathcal{A}_{ij} = 1$ if $i$ and $j$ are connected and $\mathcal{A}_{ij} = 0$ otherwise. We restrict our attention to the dynamics of \eqn\ref{eqn:DRDN} on star graphs with $k$ arm nodes, illustrated in \fig\ref{fig:Dynamics}.

We additionally employ the method of phase-reduction to create a further simplified model of the phase-locked dynamics \cite{BZSTARsupp,Izhikevich:Systems, Monga2018}. The approach assumes that the state of an isolated multi-variable chemical oscillator with limit-cycle dynamics can be fully described with a single phase variable. For weakly-coupled oscillators, the slow dynamics of the network, obtained by averaging over one oscillation period, take the form $\dot{\Phi}_i=\omega + \sum_{j=1}^{N}{\mathcal{A}_{ij}\mathscr{H}_{ij}}\left(\Phi_{ji}\right)$ where $\Phi_i$ and $\dot{\Phi}_i$ are the phase and instantaneous frequency of the $i^\mathrm{th}$ oscillator, respectively, $\omega$ is the frequency of an isolated oscillator (assumed to be identical for all oscillators), $\Phi_{ji}=\Phi_j-\Phi_i$ is the phase difference between two oscillators, and $\mathscr{H}_{ij}$ is a numerically calculated interaction function describing how the instantaneous frequency of oscillator $i$ is changed by the presence of oscillator $j$, derived from  \eqn\ref{eqn:DRDN} and shown in \fig\ref{fig:phasemodel}~\footnote{We note that by convention\cite{Izhikevich:Systems} $\mathscr{H}_{ij}$ is defined to be a function of $\Phi_j - \Phi_i$; however, in \fig\ref{fig:phasemodel} we plot as a function of $\Phi_i - \Phi_j$ to compare to $\mathscr{G}$. This change in variables mirrors the function about $\Phi=\pi$, changing the sign of the slope $\frac{d\mathscr{H}_{ij}}{d\Phi_{ji}}=-\frac{d\mathscr{H}_{ij}}{d\Phi_{ij}}$.}. We model oscillator interactions as occurring entirely through diffusive transport of the inhibitory species bromine; oscillator $i$ is delayed by oscillator $j$ for nearly all $\Phi_{ij}$, with the the largest delay occurring just before oscillator $i$ is about to undergo a transition from the reduced to oxidized state. Experimental conditions are chosen to be consistent with the model \cite{NingPCCP:2014,Ning:2014}.

\begin{figure}
\begin{center}
\includegraphics[width=1.0\columnwidth]{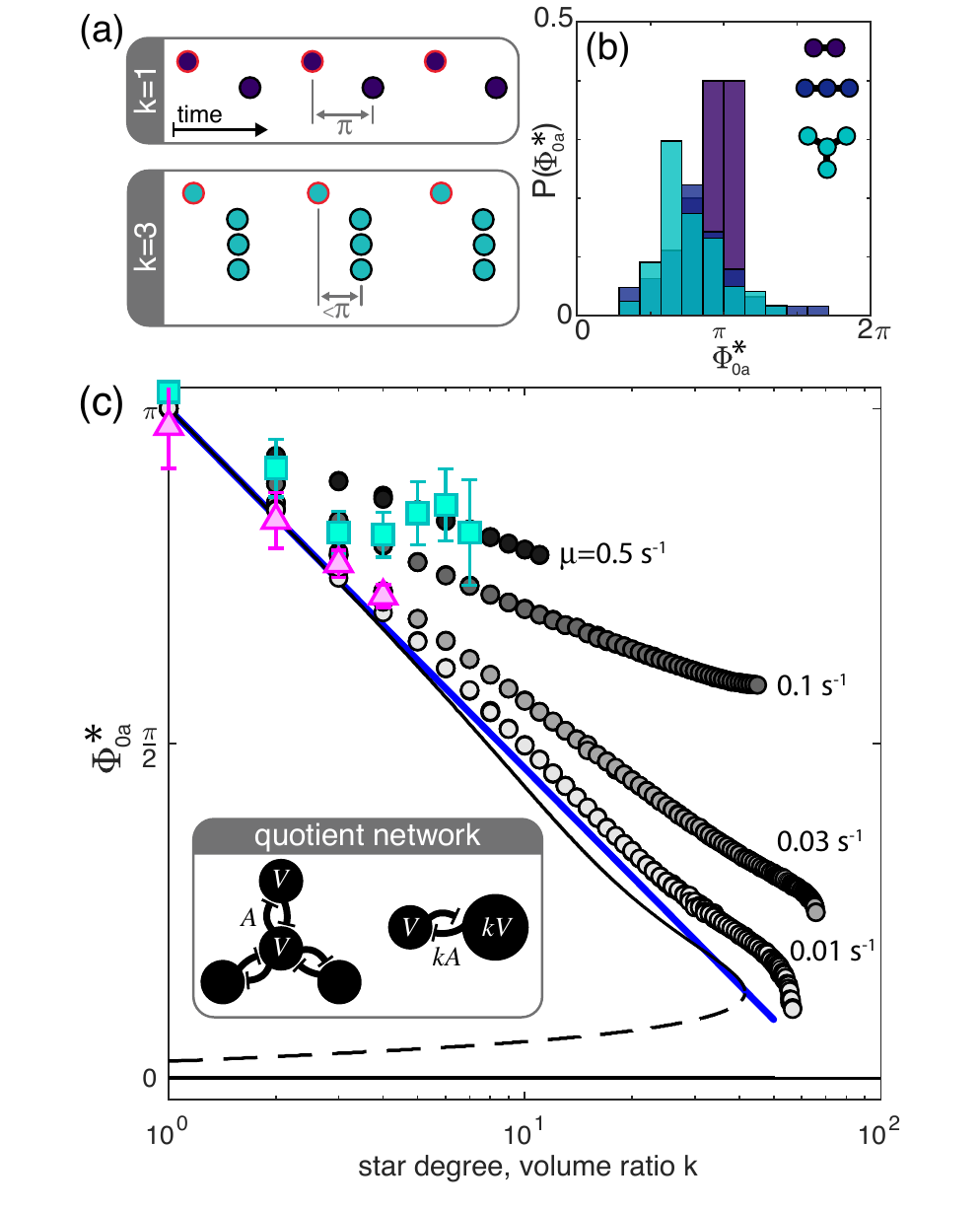}
\caption{\textbf{Locking angle $\Phi_{0a}^*$ depends on star degree.} \textbf{(a)} Illustration showing how additional arm nodes impact locking angle, \textbf{(b)} Histogram of experimental results for $k=1, 2, 3$ and all coupling strengths, \textbf{(c)} Bifurcation diagram comparing phase model (solid black lines are stable fixed points, dashed are unstable), point model (\eqn \ref{eqn:DRDN}) with heterogeneity in acid concentration, $\mu=$ 0.066~s$^{-1}$ (squares), point model with identical oscillators, $\mu=$ 0.01-0.5~s$^{-1}$ (circles), and experiment (triangles). Error bars show standard error. For $\log{k}\ll1$, $\Phi_{0a}^*$ is given by \eqn\ref{eq:logk} (blue  line).}
\label{fig:lockingangle}
\end{center}
\end{figure}

\textbf{State Diagram.} We observed three dynamical attractors as a function of the number of arms in a star graph and the coupling strength (\fig\ref{fig:phasediagram}). The  states are (\textit{i}) \textit{unlocked} (movie S5); unsynchronized oscillations of all nodes, (\textit{ii}) \textit{locked} (movies S2-S4); arm-hubs synchronized in-phase with a $k$-dependent phase shift between the arm and central hub, and (\textit{iii})  \textit{center silent} (movie S6);  a non-oscillating, or intermittently oscillating central hub and unsynchronized oscillations of all arm-hubs. Using photochemical inhibition, we are also able to change state dynamically by manipulating topology. Shining strong light on a node inhibits oscillation, effectively pruning that node from the network. We were able to induce a transition from center silent to locked in a 5-arm star by shining light on two arms, effectively transforming the network to a 3-arm star while leaving the coupling constant unchanged. We also induced a transition from locked to unlocked by shining light on the hub of a 3-arm star, as shown in movies S7-S8 and \fig S2 \cite{BZSTARsupp}.

In \fig\ref{fig:phasediagram}, experiment and theory are compared.  When the coupling strength is low (large drop separation), the unlocked state is observed. A network is considered unlocked if a steady-state locking angle is not achieved during the experiment, for theory, we examine a time window commensurate with experiments corresponding to $\sim$20 oscillations. For moderate star degree, as one increases the coupling strength,  phase locking is observed. Further increases to the coupling strength results in  center-silent dynamics. For large star degree unlocking proceeds directly to center-silent. Conversely, as star degree is lowered, the coupling strength range over which phase-locking occurs broadens. An analytic expression for the diffusive rate $\mu~[\mathrm{s}^{-1}]$ is presented in the appendix, we expect $\mu$ to be inversely proportional to the drop separation and to the volume of the receiving drop \cite{BZSTARsupp}.

In order to match the theory-predicted state diagram with the experiment we introduced normally distributed variations of 7.5\% in H$^+$ which created a distribution in oscillator frequency. We ran each parameter combination ($k$ and $\mu$) twenty times, resampling the variation in chemistry and initial conditions each time. We also introduced a coupling strength reduction (fudge) factor $f=0.15$ to modify the predicted $\mu$. Heterogeneity is needed to both destroy synchrony when coupling is weak \cite{Strogatz:Chaos} and move the transition from locked to center-silent to lower $k$. Phase diagrams showing theoretical predictions without these modifications are shown in the supplement, \fig S4 \cite{BZSTARsupp}.

Variations in parameters are justified because the inner, aqueous phase is composed of two reactant streams; fluctuations in flow rates can therefore produce droplets of varying composition and size. For simplicity, we consider chemical heterogeneity alone.  The reduction of the effective coupling coefficient has been noted previously \cite{Tompkins:2014}.  These results indicate our idealized model is qualitatively, but not quantitatively, correct and further, that system heterogeneity is an important parameter.
 
\textbf{Steady State Locking Angles and Phase Model.} The phase interaction function $\mathscr{H}_{ij}$, assuming only bromine transport between oscillators, is nearly completely negative meaning that coupled oscillators mutually delay one-another, \fig\ref{fig:phasemodel} (bold curve). This produces a steady state phase shift between two oscillators, at the maximum phase difference of $\pi$, \fig\ref{fig:phasemodel} ($k=1$ curve). Within the phase-locked regime, we experimentally observe that as the number of arms increases, the delay between when the hub and arm nodes oxidize decreases, see color-coded time traces in \fig\ref{fig:Dynamics}d \& e and \fig\ref{fig:lockingangle}. We examine in detail this dependence using the numerically constructed phase model.


Since all solutions exhibit arm-locked dynamics, a consequence of the arms forming a single orbit of the star network's graph \cite{Pecora2014b}, we examine the dynamics along the arm-synchronized manifold $\Phi_1=\Phi_2= \dots = \Phi_k$ and reduce the system to a single degree of freedom $\Phi_{0a}=\Phi_0-\Phi_a$ with arm-synchronized cluster $a$ and hub $0$. The dynamics of this quotient network are described by 
\begin{equation}\dot{\Phi}_{0a}=k \mathscr{H}_{0a}\left(-\Phi_{0a}\right)-\mathscr{H}_{a0}\left(\Phi_{0a}\right)=\mathscr{G}\left(\Phi_{0a}\right).
\label{eqn:coarsephasemodel}
\end{equation}
Examining arm-synchronized dynamics of $N$ oscillators is equivalent to considering a heterogeneous pair: a node containing a single unit of volume $V$ and a large node of volume $kV$ diffusively coupled through a conduit with $k$-times the cross sectional area $A$ of the connections in the original network, inset of \fig\ref{fig:lockingangle}c. This volume difference manifests as an asymmetric coupling constant  $\mu_{0a}=k\mu_{a0}$. We report the system dynamics for various $k$ rescaled by the coupling strength and star degree $\mathscr{G}/\left(k\mu\right)$ so that as $k$ becomes large the plotted amplitude of the dynamics does not grow as well, \fig\ref{fig:phasemodel}B.

The system's locking angles are readily given by the fixed points $\mathscr{G}\left(\Phi_{0a}^*\right)=0$ with stability determined by the slope $\mathscr{G}'$: if $\mathscr{G}'<0$, the point is stable; if $\mathscr{G}'>0$, unstable. The results show that for two equally sized nodes, there are four fixed points, two of which are stable $\Phi_{0a}^*=0,\pi$.  However, the basin of attraction associated with $\Phi_{0a}^*=0$ is so small, it isn't visible in \fig\ref{fig:phasemodel}B. In contrast, the basin for $\Phi_{0a}^*=\pi$ is larger and deeper and therefore more accessible and robust against differences in the intrinsic frequencies of the wells, as seen experimentally~\cite{Epstein:2010,Epstein:2011}. 

The phase model predicts that as the volume ratio increases from unity, the hub oscillator is more delayed by the collective action of the arm cluster, shifting $\Phi_{0a}^*$, \fig\ref{fig:lockingangle}. For example, for $k=3$, theory predicts the locking angle is reduced to $0.76 \pi$. As the volume ratio increases, the fixed point continues to move until the volume ratio reaches 1:33.8 where a saddle-node bifurcation eliminates the attractor, leaving $\Phi_{0a}^*=0$ as the only attractor, shown as a black line in \fig\ref{fig:lockingangle}.

These predictions compare favorably to experiment,
and the full chemical model without heterogeneity, \fig\ref{fig:lockingangle}. In the limit of weak coupling strength, assumed by the creation of the phase model, the locking angle is $\mu$-independent. As coupling strength is increased in the point model, predictions diverge with $k$ more rapidly. Additionally, for strong coupling ($\mu=0.1 - 0.5$ s$^{-1}$) the system transitions to center-silent while for weaker coupling ($\mu=0.01 - 0.03$  s$^{-1}$) the system instead transitions to in-phase synchrony as predicted by the phase model. Heterogeneity introduced into the model \eqn\ref{eqn:DRDN} to reproduce the phase boundaries in \fig\ref{fig:phasediagram} does not change the overall dependence of locking angle on star degree,  illustrated with squares in \fig\ref{fig:lockingangle}.

In the limit of weak coupling, $\Phi_{0a}^*-\pi\propto-\log{k}$, as shown in \fig\ref{fig:lockingangle}. While physically $k \in \mathbb{N}$ for our system, more generally, $k$ represents the volume ratio between two diffusively-coupled oscillators and can take on any positive value. We anticipate logarithmic scaling because the deviation from $\pi$ should change sign, but not magnitude upon relabelling the quotient graph, which is equivalent to inverting the volume ratio $\log{(k)} = -\log{(1/k)}$. To identify the pre-factor, we perform a regular perturbation expansion of the locking angle with $\log\left(k\right)$ as the expansion variable, with details provided in the supplement \cite{BZSTARsupp}. The expansion approximates the location of the attractor of \eqn\ref{eqn:coarsephasemodel} according to the scaling law
\begin{equation}
\Phi_{0a}^*=\pi+\log\left(k\right)\frac{1}{2}\mathscr{H}_{0a}\left(\Phi_{a0}\right)\left.\left(\frac{d\mathscr{H}_{0a} }{d\Phi_{a0}}\right)^{-1}\right|_{\Phi_{a0}=\pi},
\label{eq:logk}
\end{equation}
which compares well to the phase model up to $\log{k}\sim\mathscr{O}\left(1\right)$. The first term, $\Phi_{0a}^{*\left(0\right)}=\pi$, arises from symmetry and is agnostic to oscillator chemistry and network topology. In contrast, the next order correction encodes information specific to both. Information about the chemical reactions enters through the oscillator's phase response curve and coupling to adjacent oscillators through $\mathscr{H}$, while information about network topology enters through $\log{k}$. We note that while networks of repulsively coupled oscillators have been modeled simply as negatively coupled Kuramoto oscillators $\dot{\Phi}_i=\sum_{j=1}^N \mathcal{A}_{ij}\sin{\left( \Phi_j-\Phi_i\right)}$ \cite{Acebron2005,Giver2011a}, our result shows explicitly that because $\mathscr{H}\left(\pi\right)=\sin\left(\pi\right)=0$, sine coupling does not predict the observed topology-dependence for branching networks.

\textbf{Discussion and Conclusion.} 
In this Letter we show how the coupling strength and topology of an inhibitor-coupled BZ reaction-diffusion star-network controls transitions between distinct dynamic states.  We demonstrate that the high-dimensional reaction-diffusion system of the star network, consisting of $(k+1)\times M$ variables (\eqn\ref{eqn:DRDN}), can be reduced to a low-dimensional phase model of $k$ variables and further simplified to a quotient graph of only 1 phase variable that remains able to make quantitative predictions of the locking angle, \eqn\ref{eq:logk}. The symmetry-based argument in reducing the star network to a quotient graph (\eqn\ref{eqn:coarsephasemodel}) requires perfectly identical oscillators. The strong agreement with experiment indicates that topology-dependent dynamics are robust against experimental imperfections. One of the ongoing objectives of the study of networks is to elucidate the connection between topology and dynamics. Our theoretical prediction of the locking angle,  \eqn\ref{eq:logk}, explicitly separates topological effects from chemical dynamics. The theory is universal and therefore readily transferable to other oscillator networks. These results demonstrate the utility of model experimental reaction-diffusion systems for both testing theories of network dynamics and providing engineering principles for dynamic soft materials.


\begin{acknowledgments}
\textbf{Acknowledgments.} N.T, M.C.C. M.M.N, B.B., and J.H. acknowledge financial support from NSF DMREF-1534890, and U. S. Army Research Laboratory and the U. S. Army Research Office under contract/ grant number W911NF-16-1-0094. The microfluidic experiments were developed by N.T. through NSF MRSEC DMR-1420382. We would also like to thank Michael Rubenstein for his thoughtful discussions on the locking angle. 
\end{acknowledgments}

\bibliographystyle{apsrev4-1}

%

\end{document}